# Thermoelastic equation of state and melting of Mg metal at high pressure and high temperature

Alexandre Courac,[*,a] Yann Le Godec,[a] Vladimir L. Solozhenko,[b] Nicolas Guignot,[c] and Wilson A. Crichton[d]

[a] *Institut de Minéralogie, de Physique des Matériaux et de Cosmochimie (IMPMC), Sorbonne Université, UMR CNRS 7590, Muséum National d'Histoire Naturelle, IRD UMR 206, 75005 Paris, France*

[b] *LPMTM-CNRS, Université Paris Nord, 93430 Villetaneuse, France*

[c] *Synchrotron SOLEIL, 91192 Gif-sur-Yvette, France*

[d] *ESRF - The European Synchrotron, 71, avenue des Martyrs, 38000 Grenoble, France*

**Abstract**

The *p-V-T* equation of state of magnesium metal has been measured up to 20 GPa and 1500 K using both multianvil and opposite anvil techniques combined with synchrotron X-ray diffraction. To fit the experimental data, the model of Anderson-Grüneisen has been used with fixed parameter $\delta_T$. The 300-K bulk modulus of $B_0 = 32.5(1)$ GPa and its first pressure derivative, $B_0' = 3.73(2)$, have been obtained by fitting available data up to 20 GPa to Murnaghan equation of state. Thermal expansion at ambient pressure has been described using second order polynomial with coefficients $a = 25(2) \times 10^{-6}$ K$^{-1}$ and $b = 9.4(4) \times 10^{-9}$ K$^{-2}$. The parameter describing simultaneous pressure and temperature impact on thermal expansion coefficient (and, therefore, volume) is $\delta_T = 1.5(5)$. The good agreement between fitted and experimental isobars has been achieved to relative volumes of 0.75. The Mg melting observed by X-ray diffraction and *in situ* electrical resistivity measurements confirms previous results and additionally confirms the *p-T* estimations in the vicinity of melting.

**Introduction**

Study of magnesium metal under high pressure is interesting not only from fundamental point of view for searching new exotic matter state at extremely high pressures,[1] but also for understanding of thermochemical properties of Mg-bearing systems for synthesis of diamond[2] and energetic materials[3, 4] at high pressure – high temperature (HPHT) conditions. The phase transformations in Mg have been studied at HPHT by Errandonea et al.[5] and Stinton et al.[6] Above 1200 K in the 10-20 GPa range, the formation of a double-hexagonal close-packed (dhcp) phase has been observed from initial hcp structure. The dhcp phase may be recovered as metastable at room temperature and high pressure in a diamond anvil cell (at ~8 GPa). At the same time, data[5, 6] analysis shows that the volume change during hcp to dhcp transformation, if exists, is close to zero.

Magnesium compressibility and thermal expansion have been previously measured by both *in situ* X-ray diffraction and dilatometry.[5-10] The most recent and accurate results give the bulk modulus $B_0 = 32.5(4)$ GPa with its pressure derivative $B'_0 = 4.05(5)$.[6] The linear thermal expansion ($L(T)$ at 0.1 MPa) of Mg up to the melting temperature well follows the square polynomial with thermal expansion coefficient $\alpha_L = 25 \times 10^{-6}$ K$^{-1}$ and its first temperature derivative $\alpha'_L = 18.8 \times 10^{-9}$ K$^{-2}$ (at 273 K).[11] The slightly non-isotropic lattice expansion has been reported in ref.[9].

Previously the *p-V-T* equation of state (EoS) was only reported in ref.[5] up to 19 GPa and melting temperatures, however, the proposed analytical expression for the *p-V-T* EoS suffers from thermodynamic inconsistency (e.g. crossing isotherms) due to the linear extrapolation of temperature dependence of the bulk modulus. At the same time, the relatively small number of data was not sufficient for fitting of the parameters describing thermal dependencies of bulk modulus and thermal expansion. The thermodynamically consistent model with smaller number of assumptions and fitting parameters is required to treat such data, for example, the model[12-14] with constant Anderson-Grüneisen coefficient $\delta_T$.[15, 16]

In the present work, the *p-V-T* equation of state of Mg has been studied by *in situ* synchrotron X-ray diffraction (XRD) up to 20 GPa and melting temperatures. The analytical expression for the *p-V-T* equation of state combines previously reported *p-V* and *V-T* data (at 300 K and 0.1 MPa, respectively) and one fitting parameter, the Anderson-Grüneisen coefficient $\delta_T$.[15, 16]

## 2. Experimental

Chemically pure Mg powder (Alfa Aesar, 99.99 at%) has been used as a starting material. The experiments have been conducted in MgO capsules with various pressure media.

The Paris–Edinburgh (PE) press at beamline PSICHÉ of synchrotron SOLEIL was used for obtaining *in situ* data at high temperatures in the 5.0-6.5 GPa range. Opposite anvils with standard boron-epoxy gasket (pressure medium) were employed to pressurize the sample. A tubular graphite resistive furnace allowed heating up to 1500 K under high pressure. The Mg-C mixtures were loaded into an MgO capsule in order to isolate the sample from the heater. Pressure and temperature estimations were made using 300-K equations of states of Mg[6] and MgO[17] and temperature calibration curve obtained using Si melting point at HPHT.[18, 19] The phase transformations were observed by energy-dispersive X-ray diffraction ($2\theta = 8.0(2)°$). The system was calibrated using an Au standard.

The high-pressure experiments at 1.5-8 GPa (according to the hBN equation of state)[20] with direct temperature measurements were carried out using the multianvil X-ray system MAX80[21] with anvils of tungsten carbide. The diffraction measurements were performed in energy-dispersive mode at beamline F2.1 (HASYLAB-DESY, Hamburg). The Mg + B (amorphous) and Mg + hBN samples were prepared in a glove box with dry argon. X-ray patterns were collected on a Canberra solid state Ge-detector with fixed Bragg angle $2\theta = 9.96(2)°$.[22] The details of experiments and high-pressure setup have been described elsewhere. The temperature of the high-pressure cell was controlled by a Eurotherm PID regulator within 2 K. The sample temperature was measured by a Pt 10% Rh-Pt thermocouple with its junction 300 µm below the sample region under study. The primary polychromatic synchrotron beam collimated down to 60×100 µm² was perpendicular to the vertical axis of the sample chamber. The energy-dispersive diffraction patterns were collected *in sit*u at a constant pressure in an "auto-sequence" mode in the course of a linear heating (or cooling) at a rate of 30 K min$^{-1}$; the time of data collection for each pattern was 30 s.

*In situ* experiments at 8-20 GPa (Mg + glassy carbon) were performed using the 20MN Voggenreiter press at beamline ID06 of the ESRF.[23] The Mg + C mixture was ground in a ceramic mortar inside a high-purity Ar glovebox and loaded into an MgO capsule. The sample was then introduced (under Ar atmosphere) into a 10/5 multianvil assembly (MgO:Cr$_2$O$_3$ octahedron with 10 mm side compressed by eight WC cubic anvils with 5mm-side triangular truncations), equipped with graphite or Re furnaces. Temperatures and pressures were monitored using the 300-K equations of state of MgO[17] and Mg[6], in parallel with estimation from previous calibration curves (Si EoS and melting)[18, 19] and W-Re thermocouples. Monochromatic X-ray diffraction data were taken using the 0.3757 Å (33 keV). The beam was collimated to define a horizontal beam size of ~1 mm to ensure that the whole sample was probed. Diffraction patterns were collected on an azimuthally-scanning Detection Technology X-Scan c series GOS linear detector (the data collected continuously, at fixed azimuth).

The absence of remarkable broadening has been used as criterion of reaching the quasi-hydrostatic conditions and was typically observed above 600 K. Only these parts of experimental quasi-isobars have been used for fit to the equation of state. (In practice, pressure changes during heating, and the sample is not at the pressure estimated before heating). The $p,T$-values were refined using cross-calibration (MgO, hBN) or a pressure gauge compound with thermocouple. The data on thermal expansion of magnesium at given pressure was in some cases limited to the onset of reaction with pressure medium (typically at ~1500 K for Mg + C mixture[4] and ~1400 K for Mg + B mixture[22] in the pressure range under study). The reaction(s) at high temperatures and/or melting precluded collection of data on cooling and upon decompression.

In order to avoid the possible systematic errors in the estimation of melting points by XRD (recrystallization and single crystal growth may also cause the apparent disappearance of diffraction rings) at various pressures, we also monitored the onset of melting via electrical measurements of the furnace assembly (Fig 1). In fact, when using the solid Mg + glassy C mixture in contact with graphite heater, the resistance of heater remains very close to that of heater alone. On increasing temperature, a trace dominated by a typical semiconductor furnace is evident (Fig 1b). At melting, Mg penetrates the whole assembly and causes a significant drop in resistance enabling us to pinpoint the onset clearly. Similar method has been recently used for melting curve measurement of $Na_4Si_4$.[24] The knowledge of Mg melting curve also allowed to refine the pressure or temperature values in some cases.

## 3. Computational method

For the equation of state data fitting we have used integrated form of the Anderson-Grüneisen equation.[15, 16] In previous works[12-14] we have shown that this equation, which takes into account the pressure dependence of thermal expansion coefficient $\alpha$ only through the volume change, i.e.

$$\alpha(p,T) = \alpha(0,T)\left[\frac{V(p,T)}{V(0,T)}\right]^{\delta_T}, \qquad (1)$$

can be integrated (under the assumption that $\delta_T$ is constant over the studied *p-T* domain) to

$$V(p,T) = \left[V(0,T)^{-\delta_T} + V(p,300)^{-\delta_T} - V(0,300)^{-\delta_T}\right]^{-1/\delta_T}, \qquad (2)$$

where thermal expansion (i.e. $V(0,T)$ at 0.1 MPa) and isothermal compression (i.e. $V(p,300)$ at 300 K) can be presented in any analytical form, e.g. polynomial

$$V(0,T) = V(0,300)\,[1 + a\,(T-273) + b\,(T-273)^2 - a\,(300-273) + b\,(300-273)^2]^3 \qquad (3)$$

with $a = \alpha_L$ and $b = 0.5 \times \alpha'_L$ and Murnaghan (or any other) equation of state[7]

$$V(p,300) = V(0,300)\left(1 + B'_0\,p/B_0\right)^{-1/B'_0}. \qquad (4)$$

Thus, a set of parameters needed to describe an EOS using equations (2-4) is $V_0 \equiv V(0,300) \equiv M/\rho_0$, $B_0$, $B'_0$, $a$, $b$ and $\delta_T$.[13] Such form of the EOS, (2), allows one easily approximate the $V(p,T)$ in the domain of interest for various objectives such as *p-T* refinement or phase equilibria calculation.[2, 12-14, 25]

## 4. Results and Discussion

At ambient conditions Mg has hexagonal structure [space group *P6₃/mmc* (No. 194)] with unit cell parameters $a = 3.209(2)$ Å and $c = 5.211(3)$ Å. The synchrotron X-ray diffraction patterns of Mg show three lines, i.e. *100*, *002* and *101*, that were used to establish the molar volume $V(p,T)$ at HPHT conditions. The *in situ* XRD data on thermal expansion at given pressure (quasi-isobars) and observation of melting (disappearance of solid Mg reflections, appearance of characteristic halo of liquid) are shown at Fig. 2.

The *p-V-T* data (Tabs. 1 and 2) of our experiments and from refs.[5, 6] are plotted at Fig. 3a (symbols) as compared with calculated isobars. Each color (symbols correspond to experimental points) cover the possible relative volume values $V/V_0$ at different temperatures for a 2-GPa pressure range (ten in total, covering total pressure domain of available data from 0.1 MPa to 20 GPa). The pressure change during heating was not significant above 600 K, but can be up to 2 GPa higher or lower than before heating ("cold compression").

The best fit of experimental data has been obtained using the 300-K equation of state of Stinton et al.[6] (coefficients for Eq. 4 are $B_0$ = 32.5(2) GPa, $B'_0$ = 3.73(2), reproducing the data[6] up to 25 GPa to high accuracy of relative volume, i.e. with uncertainty $\delta(V/V_0) < 10^{-3}$). The thermal expansion has been taken into account by the analytical expression, valid up to melting temperature (Eq. 3 with parameters $a = 25 \times 10^{-6}$ K$^{-1}$ and $b = 9.4 \times 10^{-9}$ K$^{-2}$). As a guide for eyes to judge the quality of fit, we use the color match between theoretical domains and symbols at Fig. 3a. The parameter $\delta_T$ = 1.5(5) gives the best fit for the $p$-$V$-$T$ experimental data. Isothermal Anderson Grueneisen parameter $\delta_T$ for Mg has been previously evaluated by two alternative methods,[26] $\delta_T$ = 1.66 using experimental data on the $T$-dependence of elastic constants[27] and $\delta_T$ = 2.69 from theoretical lattice dynamic study.[28] One can see reasonable agreement of our $\delta_T$ value obtained by the EoS fit with elastic measurements.[26, 27] The values of $\delta_T$ = 1.0 and 2.0 also gives reasonable agreement of all available experimental data (Tabs. 1 and 2). Figure 4 shows the set of isobars (from 0.1 MPa up to 20 GPa) for $\delta_T$ = 1.0, 1.5 and 2.0. Significant discrepancy is observed only for very high temperatures, where solid Mg does not exist.

The *in situ* observations of melting at different synchrotron facilities (Fig. 2) and by resistance measurements (Fig. 1b) are presented at Fig. 3b. Our experimental value of the zero-pressure melting slope is d$T$/d$p$ = 60(5) K GPa$^{-1}$, in agreement with previous resistivity measurements, of 60(2) K GPa$^{-1}$.[29] At higher pressure the melting slope decrease down to 38(4) K GPa$^{-1}$ at ~10 GPa according to our estimations from the data reported in ref. [29]. Such noticeable decrease is often due to the higher compressibility of liquid phase as compared to solid.

The melting heat of Mg is $\Delta H_0$ = 8.5 kJ mol$^{-1}$,[30] and liquid density at melting temperature and ambient pressure of $\rho_m$ = 1.590(1) g cm$^{-3}$ with linear negative thermal coefficient -d$\rho_m$/d$T$ = 2.647×10$^{-4}$ g cm$^{-3}$ K$^{-1}$ up to boiling temperature at 1390 K.[31] The room-temperature density at ambient pressure of solid Mg is $\rho_0$ =1.737 g cm$^{-3}$, which gives the density estimation of 1.639 g cm$^{-3}$ at melting point (923 K) according to our EoS. The value of melting volume is thus $\Delta V_0$ = + 0.456(9) cm$^3$ mol$^{-1}$, and the estimate for the melting curve slope is d$T$/d$p$ = 47(2) K GPa$^{-1}$, which is in satisfactory agreement (22 % accuracy) with observed value of 60 K GPa$^{-1}$.[23] Alternative estimates of Mg liquid density[32] ($\rho_m$ = 1.584 g cm$^{-3}$, -d$\rho_m$/d$T$ = 2.34×10$^{-4}$ g cm$^{-3}$ K$^{-1}$) give better agreement: volume is $\Delta V_0$ = + 0.51(2) cm$^3$ mol$^{-1}$, and the estimate for the melting curve slope is d$T$/d$p$ = 53(2) K GPa$^{-1}$, i.e. only 11% below the experimental estimate.

## 5. Conclusions

Finally, the *p-V-T* equation of state of metallic Mg have been studied up to 20 GPa and 1500 K using different high-pressure apparatuses and synchrotron radiation facilities. The model describing analytical expression for the temperature and pressure dependencies of volume is consistent with EoS parameters previously derived from both 300 K and 0.1 MPa measurements, and the constant value of Anderson-Grüeneisen parameter $\delta_T$ = 1.5(5) is appropriate to describe *V*(*p,T*) all over the *p-T* range under study (to compression of 0.75). The melting *p-T* data by both XRD and electrical measurements is consistent with previous reports.

**Acknowledgements.**

OOK and YLG thank Agence Nationale de Recherche for financial support (project ANR-17-CE08-0038). We acknowledge synchrotrons ESRF, DESY and SOLEIL for provision of synchrotron radiation facilities. The *in situ* XRD experiments were performed (1) on beamline ID06-LVP at the European Synchrotron Radiation Facility (ESRF), Grenoble, France (allocated beamtime CH-5431); (2) on F2.1 beamline at HASYLAB-DESY, Hamburg, Germany; and (3) on beamline PSYCHÉ at SOLEIL, Gif-sur-Yvette, France. We are grateful to Dr. K. Spector at the ESRF for providing assistance in using beamline ID06-LVP. The authors also acknowledge Dr. H. Moutaabbid and Mr. I. Touloupas for assistance in high pressure experiments at IMPMC.

**Table 1**. Pressure $p$, temperature $T$ and relative volume $V/V_0$ data ($V_0 = 13.98$ cm$^3$g$^{-1}$) from previous reports.[5, 6]

| $p$, GPa | $T$, K | $V/V_0$ | $p$, GPa | $T$, K | $V/V_0$ |
|---|---|---|---|---|---|
| Data from ref. [5] | | | | | |
| 0 | 300 | 0.9957 | 12.07 | 1077 | 0.8483 |
| 0 | 300 | 0.9978 | 12.07 | 1127 | 0.8519 |
| 0 | 300 | 1.0050 | 12.07 | 1127 | 0.8562 |
| 0 | 300 | 0.9835 | 12.08 | 977 | 0.8154 |
| 0.65 | 300 | 0.9692 | 12.4 | 927 | 0.8097 |
| 3.77 | 300 | 0.8719 | 12.4 | 927 | 0.8097 |
| 4.53 | 300 | 0.8927 | 12.5 | 1377 | 0.8218 |
| 4.98 | 300 | 0.9105 | 12.82 | 877 | 0.8104 |
| 5.25 | 300 | 0.8898 | 13.1 | 1247 | 0.8233 |
| 7.02 | 877 | 0.9341 | 13.65 | 677 | 0.8068 |
| 8.05 | 300 | 0.8447 | 13.7 | 1177 | 0.7954 |
| 8.6 | 1127 | 0.9206 | 14.08 | 1047 | 0.7546 |
| 9.64 | 1277 | 0.8826 | 14.38 | 477 | 0.7939 |
| 10.5 | 1477 | 0.8927 | 14.6 | 1047 | 0.7675 |
| 10.7 | 1427 | 0.8898 | 14.75 | 300 | 0.7811 |
| 11.1 | 1377 | 0.8798 | 15.3 | 877 | 0.7875 |
| 11.3 | 1327 | 0.8733 | 15.8 | 677 | 0.7789 |
| 11.3 | 1327 | 0.8691 | 16.3 | 477 | 0.7646 |
| 11.56 | 1277 | 0.8619 | 16.8 | 300 | 0.7560 |
| 11.8 | 1227 | 0.8633 | 17.65 | 300 | 0.7525 |
| 11.8 | 1227 | 0.8612 | 18.5 | 577 | 0.7625 |
| 12.06 | 1177 | 0.8483 | 18.6 | 300 | 0.7596 |
| 12.07 | 1027 | 0.8483 | 18.5 | 577 | 0.7625 |
| 12.07 | 1077 | 0.8505 | | | |
| Data from ref. [6] The $V/V_0$ values were evaluated from Fig. 4 of ref.[6], pressure at 300 K (value in parences) was reestimated using Mg equation of state[6] and better agree with observed relative volumes of Mg and with the fact that at 1315 K Mg still remain solid.[6] | | | | | |
| 3.6 (6.5) | 300 | 0.850 | 5.0 (6.5) | 1170 | 0.898 |
| 4.4 (6.5) | 1018 | 0.866 | 5.6 (6.5) | 1315 | 0.907 |

**Table 2**. Pressure $p$, temperature $T$ and relative volume $V/V_0$ data ($V_0 = 13.98$ cm$^3$g$^{-1}$) from our experiments.

| $p$, GPa | $T$, K | $V/V_0$ | $p$, GPa | $T$, K | $V/V_0$ |
|---|---|---|---|---|---|
| HASYLAB-DESY data. | | | | | |
| 1.5 | 587 | 0.9971 | 7.0 | 775 | 0.8866 |
| 2.3 | 877 | 0.9978 | 7.0 | 675 | 0.8781 |
| 2.7 | 1047 | 0.9950 | 6.0 | 300 | 0.8697 |
| 2.7 | 977 | 0.9964 | 6.0 | 875 | 0.9123 |
| 7.0 | 300 | 0.8561 | 6.0 | 1025 | 0.9238 |
| SOLEIL data. | | | | | |
| 4.9 | 300 | 0.8871 | 5.0 | 300 | 0.8851 |
| 5.2 | 400 | 0.8848 | 5.3 | 600 | 0.8977 |
| 5.5 | 503 | 0.8893 | 5.2 | 750 | 0.9127 |
| 5.9 | 601 | 0.8904 | 5.7 | 930 | 0.9173 |
| 6.0 | 700 | 0.8932 | 5.6 | 950 | 0.9231 |
| 6.0 | 804 | 0.8988 | 5.5 | 1150 | 0.9443 |
| 6.0 | 902 | 0.9110 | 5.9 | 1250 | 0.9396 |
| 6.0 | 937 | 0.9133 | | | |
| 6.0 | 972 | 0.9167 | | | |
| 6.1 | 1007 | 0.9194 | | | |
| 6.0 | 1056 | 0.9225 | | | |
| ESRF data. | | | | | |
| 8.0 | 925 | 0.8791 | 17.0 | 300 | 0.7535 |
| 8.4 | 1070 | 0.8881 | 18.0 | 800 | 0.7623 |
| 8.8 | 1355 | 0.9075 | 18.1 | 1300 | 0.7830 |
| 9.0 | 1450 | 0.9121 | 18.5 | 1025 | 0.7708 |
| | | | 17.0 | 927 | 0.7824 |
| | | | 17.5 | 1127 | 0.7817 |

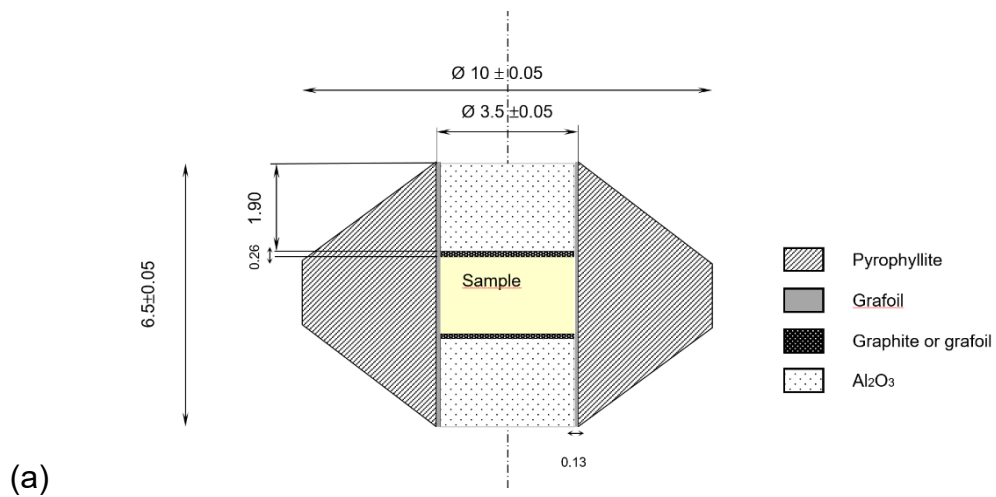

(a)

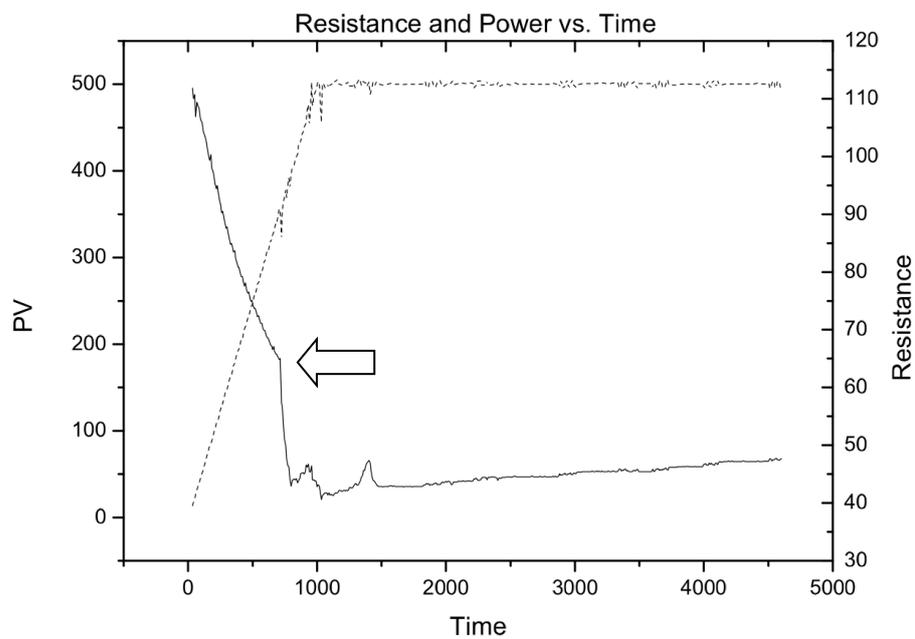

(b)

**Fig. 1 (a)** High-pressure PE cell 10/3.5 for the electrical measurements of the (sample + heater) system. **(b)** *In situ* furnace characteristics during power ramping (dotted line, left scale) of furnace containing Mg + C mixture at 5 GPa. The resistance drop (solid line, right scale) at the arrow, corresponding to 1200 K, is coincident with Mg melting.

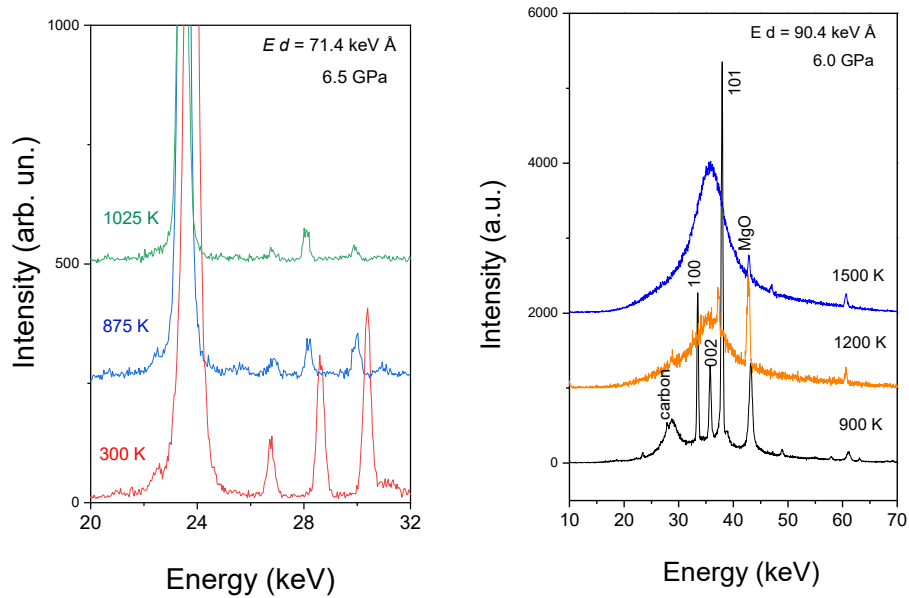

**Fig. 2** *In situ* energy-dispersive XRD data on equation of state and melting of Mg taken at HASYLAB-DESY **(a)** and at PSYCHÉ beamline at SOLEIL **(b)**.

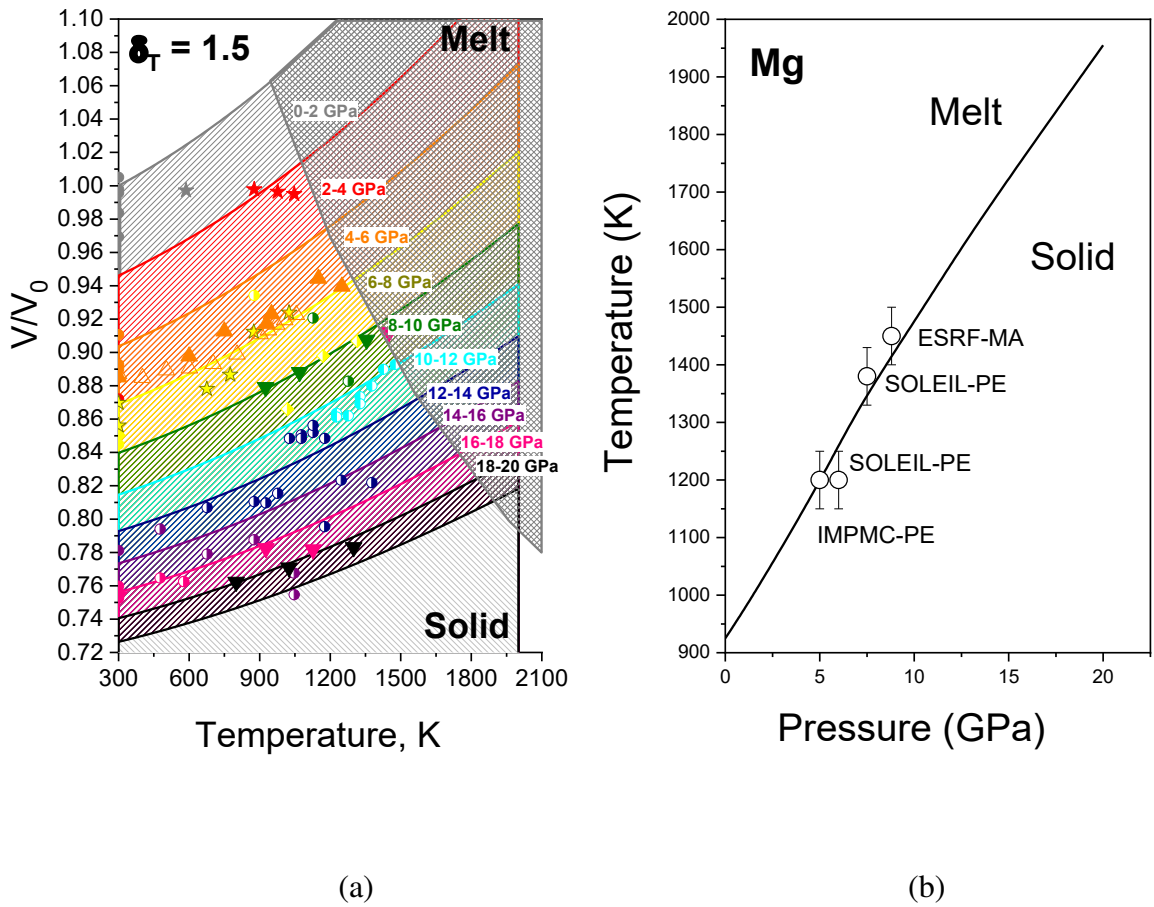

(a)                                    (b)

**Fig. 3 (a)** Theoretical isobars (2-GPa step, coloured) as compared to experimental data. Grey grid area show the domain of melted Mg (with some points of the solid Mg while coexisting with liquid). Small colored circles represent the data from refs. [5, 6] (6 and 4), the corresponding 2-GPa domains (limited by theoretical isobar curves according to Eq. 2 and melting curve) are given with the same color. Our data is presented with following symbols, depending on the synchrotron source: Β - ESRF, ξ - HASYLAB-DESY and 7,8 - SOLEIL. **(b)** Melting curve of Mg up to 20 GPa. Continuous curve corresponds to previous reports,[5, 23] while ○ symbols show melting observed by electrical resistance (IMPMC) and XRD - (SOLEIL, ESRF).

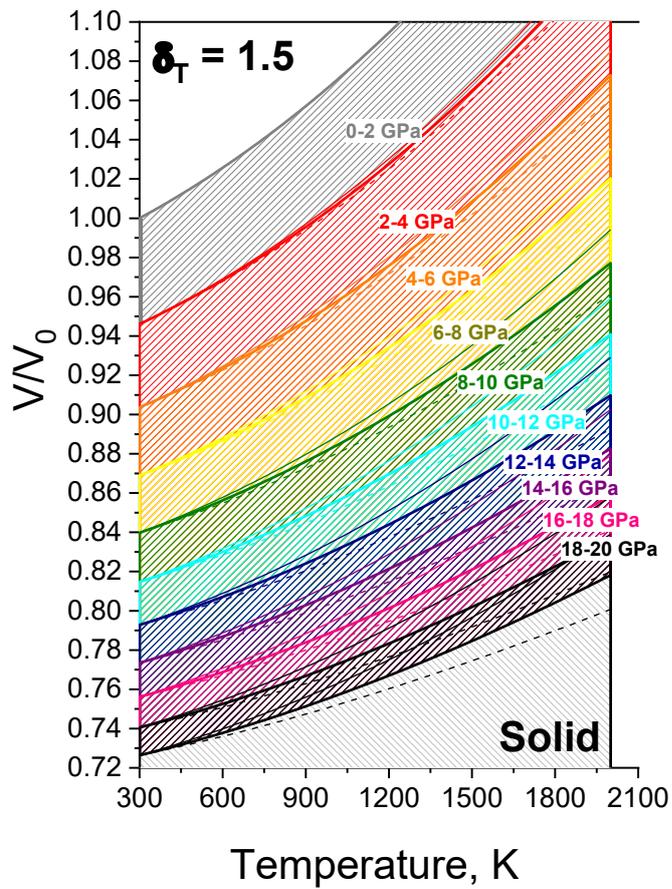

**Fig. 4** Theoretical isobars (2-GPa step, coloured) corresponding to $\delta_T = 1.5$ (thick solid lines), $\delta_T = 1.0$ (thin solid lines) and $\delta_T = 2$ (thin dashed lines).